\documentclass[10pt,twocolumn,letterpaper]{article}

\usepackage[pagenumbers]{cvpr} 

\usepackage{graphicx}
\usepackage{amsmath}
\DeclareMathOperator*{\argmin}{arg\,min}
\usepackage{amssymb}
\usepackage{booktabs}
\usepackage{subcaption}
\usepackage{float}
\usepackage{multicol}
\usepackage{multirow}

\usepackage[pagebackref,breaklinks,colorlinks]{hyperref}

\usepackage[capitalize]{cleveref}
\crefname{section}{Sec.}{Secs.}
\Crefname{section}{Section}{Sections}
\Crefname{table}{Table}{Tables}
\crefname{table}{Tab.}{Tabs.}
\newcommand\norm[1]{\lVert#1\rVert}
\def\cvprPaperID{8536} 
\def\confName{CVPR}
\def\confYear{2023}

\begin{document}

\title{MMVC: Learned Multi-Mode Video Compression with Block-based Prediction Mode Selection and Density-Adaptive Entropy Coding}
\author{Bowen Liu, Yu Chen\footnotemark[1], Rakesh Chowdary Machineni\footnotemark[1], Shiyu Liu, Hun-Seok Kim\\
University of Michigan, Ann Arbor\\
{\tt\small \{bowenliu, unchenyu, mrakeshc, shiyuliu, hunseok\}@umich.edu}
}
\maketitle

\renewcommand{\thefootnote}{\fnsymbol{footnote}}
\footnotetext[1]{Equally contributed authors.}
\begin{abstract} 
Learning-based video compression has been extensively studied over the past years, but it still has limitations in adapting to various motion patterns and entropy models. In this paper, we propose multi-mode video compression (MMVC), a block wise mode ensemble deep video compression framework that selects the optimal mode for feature domain prediction adapting to different motion patterns. Proposed multi-modes include ConvLSTM-based feature domain prediction, optical flow conditioned feature domain prediction, and feature propagation to address a wide range of cases from static scenes without apparent motions to dynamic scenes with a moving camera. We partition the feature space into blocks for temporal prediction in spatial block-based representations. For entropy coding, we consider both dense and sparse post-quantization residual blocks, and apply optional run-length coding to sparse residuals to improve the compression rate. In this sense, our method uses a dual-mode entropy coding scheme guided by a binary density map, which offers significant rate reduction surpassing the extra cost of transmitting the binary selection map. We validate our scheme with some of the most popular benchmarking datasets. Compared with state-of-the-art video compression schemes and standard codecs, our method yields better or competitive results measured with PSNR and MS-SSIM. Code is available at: \\\href{https://github.com/BowenL0218/MMVC_video_codec}{ \textit{https://github.com/BowenL0218/MMVC\_video\_codec}}
\end{abstract}

\section{Introduction}
Over the past several years, with the emergence and booming of short videos and video conferences across the world, video has become the major container of information and interaction among people on a daily basis. Consequently, we have been witnessing a vast demand increase on transmission bandwidth and storage space, together with the vibrant growth and discovery of handcrafted codecs such as AVC/H.264 \cite{6316136}, HEVC \cite{6316136}, and the recently released VVC \cite{9301847}, along with a number of learning based methods \cite{wu2018vcii, Lu_2019_CVPR, Rippel_2019_ICCV, Habibian_2019, Agustsson_2020_CVPR, Hu_2021_CVPR, Liu_2021_CVPR, yang2021hierarchical, yang2021learning, li2021deep, VCT, Li_2022_HST}.

Prior works in deep video codecs have underlined the importance of utilizing and benefiting from deep neural network models, which can exploit complex spatial-temporal correlations and have the ability of `learning' contextual and motion features. The main objective of deep video compression is to predict the next frame from previous frames or historical data, which results in the reduction of amount of residual information that needs to be encoded and transmitted. This has so far led to two directions: (1) to build efficient prediction or estimation models, and extract motion information by leveraging the temporal correlation across the frames \cite{Lu_2019_CVPR, Agustsson_2020_CVPR, yang2021hierarchical, Hu_2021_CVPR}; (2) to make accurate estimation of the distribution of residual data and push down the information entropy statistically by appropriate conditioning \cite{Habibian_2019, yang2021learning, li2021deep}. The existing works usually fall in one or a combination of the above two realms. In the light of learning capability that deep neural networks can offer, we argue and demonstrate in this work that some measures of adaptively selecting the right mode among different available models in the encoding path can be  advantageous on top of the existing schemes, especially when the adaptive model selection is applied at the block level in the feature space.
\begin{figure*}[t]
    \centering
    \includegraphics[width=0.95\linewidth]{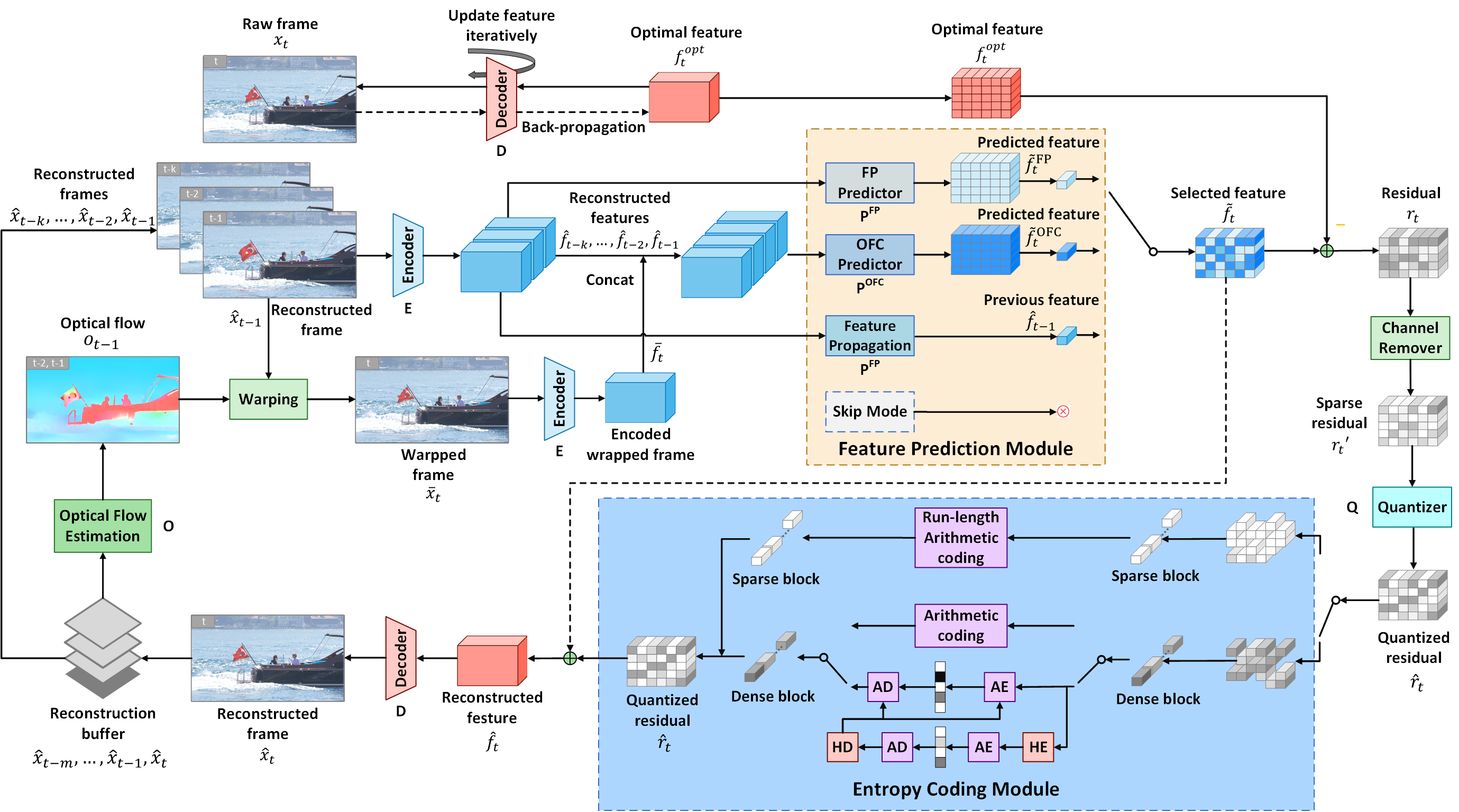}
    \vspace{-5pt}
    \caption{Overview of our proposed multi-mode video coding method. The current and previous frames are fed into the feature extractor and then go through branches of prediction modes followed by residual channel removal, quantization, and entropy coding process. We then select the optimal prediction and entropy coding schemes for each block that lead to the smallest code size.}
    \label{fig:mode}
    \vspace{-12pt}
\end{figure*}

Drawing wisdom from conventional video codec standards that typically address various types of motions (including the unchanged contextual information) in the unit of macroblocks, we present a learning-based, block wise video compression scheme that applies content-driven mode selection on the fly. Our proposed method consists of four modes targeting different scenarios: 
\begin{itemize}
    \itemsep0em 
    \item \textit{Skip} mode (S) aims to utilize the frame buffer on the decoder and find the most condensed representation to transmit unchanged blocks to achieve the best possible bitrate. This mode is particularly useful for static scenes where same backgrounds are captured by a fixed view camera.  
    \item \textit{Optical Flow Conditioned Feature Prediction} mode (OFC) leverages the temporal locality of motions. In this mode, we capture the optical flow \cite{teed2020RAFT} between the past two frames, and the warped new frame is treated as a preliminary prediction to the current frame. This warping serves as the condition to provide guidance to the temporal prediction DNN model.
    \item \textit{Feature Propagation} mode (FPG) applies to blocks where changes are detected, but there is no better prediction mode available. This mode copies the previously reconstructed feature block as the prediction, and encodes the residual from there. 
    \item For other generic cases, we propose the \textit{Feature Prediction} mode (FP) for feature domain inter-frame prediction with a ConvLSTM network to produce a predicted current frame (block).
\end{itemize}

Prior to the mode selection step, The transmitter produces the optimal low-dimensional representation of each frame using a learned encoder and decoder pair based on the image compression framework in \cite{liu_2020_BPGAN} for the mapping from pixel to feature space. The block by block difference between the previous frame and the current frame represents the block wise motion. Unlike some state-of-the-art video compression frameworks that separately encode motions and residuals, our method does not encode the motion as it is automatically generated by the prediction using the information available on both the transmitter and receiver. To adapt to different dynamics that may exist even within a single frame for different blocks, our method evaluates multiple prediction modes that are listed above at the block level. As a result, we can always obtain residuals that have the highest sparsity thereby the shortest code length per block. Furthermore, we propose a residual channel removing strategy to mask out residual channels that are inessential to frame reconstruction, exploiting favorable tradeoffs between noticeably higher compression ratio and negligible quality degradation.

Technical contributions of this work are summarized as follows: 
\begin{itemize}
    \itemsep0em 
    \item We present MMVC, a dynamic mode selection-based video compression approach that adapts to different motion and contextual patterns with \textit{Skip} mode and different feature-domain prediction paths in the unit of block. 
    \item To improve the residual sparsity without losing much quality while minimizing the bitrate, we propose a block wise channel removal scheme and a density-adaptive entropy coding strategy.
    \item We perform extensive experiments and a comparative study to showcase that MMVC exhibits superior or similar performance compared to state-of-the-art learning-based methods and conventional codecs. In the ablation study, we show the effectiveness of our scheme by quantifying the utilization of different modes that varies by video contents and scenes.
\end{itemize}

\section{Related Works}
\subsection{Learned Video Compression}
Pioneering works in learned video compression generally inherit the concept and methodology in conventional codecs. Wu \textit{et al.} \cite{wu2018vcii} propose to hierarchically interpolate the frames between a predefined interval, where both the forward and backward motions are represented by block motion vectors. DVC \cite{Lu_2019_CVPR} and Agustsson \textit{et al.} \cite{Agustsson_2020_CVPR} adopt optical flow based motion estimation and warping schemes. Habibian \textit{et al.} \cite{Habibian_2019} map a patch of frames with 3D spatial-temporal convolutions to a lower-dimensional space and make temporal predictions on the prior distribution through GRU. With the progress in designing autoencoder-based feature extraction and reconstruction as a stepping stone, recent works have achieved better performance by performing motion prediction or estimation in the feature domain (as opposed to the pixel domain), which naturally represents both motion and residual in a more information-dense form to benefit compression. FVC \cite{Hu_2021_CVPR} learns a feature space offset map as motion representation, and the motion compensation step is accomplished by deformable convolution \cite{8237351}. 

\subsection{Mode Selection and Channel Removal}
As a widely adopted tool in the conventional video coding standards, the idea of mode selection intends to evaluate different schemes on the fly to address the inter-frame temporal correlations in a context-dependant manner. Based on this concept, Ladune \textit{et al.} \cite{Ladune_2020_ModeNet} proposes a network that learns the pixel wise weighting to determine whether or not to skip encoding the respective pixel. Hu \textit{et al.} \cite{Hu_2022_CVPR} presents a hyper-prior guided mode selection scheme that compresses motion in different resolutions, and it uses a learned mask to skip the encoding of some residual features.

\subsection{Entropy Coding}
Most existing works in learned video coding adopt a learned entropy coding scheme as presented in \cite{balle2017endtoend} originally for image compression to facilitate end-to-end rate-distortion optimization. This method then evolved to provide more flexible and accurate entropy modeling by learning the distribution parameters \cite{balle2018variational, NIPS2018_8275}. Additionally for video sequences, incorporating temporal cues to obtain more accurate entropy estimation can lead to higher compression gains. RLVC \cite{yang2021learning} proposed a probability model that approximates the distribution of encoded residuals to a parameterized logistic distribution, conditioned on the feature of previous frames propagated under a recurrent setup to establish a richer temporal prior. With a similar insight, DCVC by Li \textit{et al.}  \cite{li2021deep} proposes building an entropy model with temporal conditions. Recently, Mentzer \textit{et al.} \cite{VCT} introduce a novel  transformer-based framework that establishes a temporally conditioned entropy model and abstracts all decorrelation efforts by one-shot model execution.

\subsection{Quantization and Channel Removal}
Adjusting the bin size or quantization granularity is a technique to address the uneven redundancy between channels, allowing the content of greater importance to occupy more quantized bits and the rest with less bits for the maximum quality under a similar bitrate. Cui \textit{et al.} \cite{Cui_2021_CVPR} propose scaling the residual feature with learned channel wise factors before quantization and inverse-scaling before reconstruction as an effective way of rate adaptation. In our work, to simplify the datapath but still benefit from the same concept, a channel wise binary masking (0 or 1 scaling) is applied to remove  disposable channels in the feature blocks when it offers a favorable tradeoff in the achievable rate vs. quality.

\section{Method}


We denote the original temporal sequence of raw frames as the set $X = \{x_1, x_2, \cdots, x_{t-1}, x_t,\cdots \}$. Correspondingly, on the receiver (and also on the transmitter), the reconstructed previous frames are available as $\hat{x} = \{\hat{x}_1, \hat{x}_2, \cdots, \hat{x}_{t-1}\}$. The whole system flow of our proposed video compression scheme, MMVC, is shown in Figure \ref{fig:mode}. The system can be divided into four main parts: feature extraction, temporal prediction, channel removal, and quantization plus entropy coding. We introduce and specify each part in later subsections.

\subsection{Pixel Space Preprocessing} 
As an initial step, MMVC partition two consecutive frames into $k\times k$ blocks and calculate the block wise differences to form $n = \{n_{1,2}, n_{2,3}, \cdots, n_{t-1, t}, \cdots\}$, where
\begin{equation}
n_{t-1,t}^{i,j} = \norm{x_{t-1}^{i,j} - x_{t}^{i,j}}_2^2, \;\;\;\;\; i, j \in \{1, \cdots, k\},
\end{equation}
and the superscripts $i, j$ denote the 2D position of a block. At each time step, we examine the numerical values in the block wise difference. The blocks with all-(near)zero differences are encoded using the \textit{Skip} mode, where only the block (positional) index information is recorded and transmitted. With an algorithm parameter $\epsilon\approx0$, this step generates a binary mask $m = \{m_{1,2}, m_{2,3}, \cdots, m_{t-1, t}, \cdots\}$ each of $k\times k$ elements, where 
\begin{equation}
m_{t-1,t}^{i,j} = \left\{ \begin{array}{lll}
0, & & n_{t-1,t}^{i,j} < \epsilon, \\
1, & & \text{otherwise}.
\end{array}
\right.
\end{equation}
The unchanged blocks are masked out, indicating that no prediction will be performed and no residual will be stored. Instead, they can be recovered directly from the previously reconstructed frames by copying their pixel blocks at the corresponding positions.  

To capture the pixel level motion information and utilize its temporal locality, we obtain the optical flow \cite{teed2020RAFT} $O(\cdot)$ between the previous two reconstructed frames (e.g., ${\hat{x}_{t-2},\hat{x}_{t-1}}$) and warp it to the latter one to form a set $\{\bar{x}_{2}, \cdots, \bar{x}_{t}\}$ that is commonly available on both the transmitter and the receiver. This step can be described as:
\begin{equation}
\bar{x}_t = \textit{warp}(O({\hat{x}_{t-2},\hat{x}_{t-1}}),\hat{x}_{t-1}).
\end{equation}

\subsection{Feature Extraction} 
To this point, the pixel space preparation and optical flow based warping are accomplished. To extract rich features across the frames in a compact representation, we train a set of auto-encoders together with an entropy modeling network to achieve different rate-distortion trade-off points. Given a trained encoder $E(\cdot)$ and decoder pair $D(\cdot)$, each raw frame $x_t$ is 
encoded to an optimal feature set $f^{opt}_t$ using a back-propagation based iterative scheme \cite{liu_2020_BPGAN} that refines the one-shot encoding output $f_t = E(x_t)$ through a coupled decoder $D(\cdot)$ to obtain $f^{opt}_t$ by minimizing the MSE distortion:

\begin{equation}
f^{opt}_t = \argmin_{{f}_t}d({x}_t, D({f}_t)).
\end{equation}
 Note that $f^{opt}_t$ is only available at the transmitter since the raw frame $x_t$ is not available at the receiver.
 Previously reconstructed frames $\{\hat{x}_{t-k}, \cdots, \hat{x}_{t-1}\}$, as well as the warped frame $\bar{x}_{t}$ are encoded (without iterative optimization) to corresponding feature sets $\hat{f}_t = E(\hat{x}_t)$ and $\bar{f}_t = E(\bar{x}_t)$ using the same encoder. These are commonly available on both the transmitter and receiver. 


\subsection{Feature Prediction} 
\label{method-prediction}
Fig. \ref{fig:mode} depicts our feature space prediction and mode selection strategy. Prior to the mode selection step, we assume the optimal low-dimensional representation of current frame $f^{opt}_t$ and the binary mask $m_{t-1, t}$ for \textit{Skip} mode indication are ready at the transmitter. We replace conventional motion estimation, compression, and compensation steps with feature domain prediction optionally conditioned by pixel-domain optical flow. Residual is the difference between the optimal feature $f^{opt}_t$ and the predicted feature. However, this might lead to large residuals when the prediction is not accurate. To accommodate the rich variety of motions, we introduce a prediction method consists of three mode branches, and make mode selections for \textit{each block} (not for the entire frame) based on the entropy of residuals.  

The \textit{Feature Prediction} branch is implemented with a ConvLSTM network, where the predictor $P^{\mathrm{FP}}$ takes the reconstructed features $\{\hat{f}_{t-k}, \cdots, \hat{f}_{t-1}\}$ as inputs, and  produces a predicted current frame feature representation, capturing temporal correlation in the feature domain: 
\begin{equation}
    \tilde{f}^{FP}_{t} = P^{\mathrm{FP}}(\hat{f}_{<t}), \;\; \text{with} \;\; r^{\mathrm{FP}}_t = f^{opt}_t - \tilde{f}^{\mathrm{FP}}_t.
\end{equation}

To augment this prediction process with more contextual cues from certain scenes, we form another prediction path, called \textit{Optical Flow Conditioned Feature Prediction}. It uses the optical flow warped feature $\bar{f}_t$ as a conditional input for a prediction network $P^{\mathrm{OPC}}$:
\begin{equation}
    \tilde{f}^{\mathrm{OFC}}_t = P^{\mathrm{OFC}}(\hat{f}_{<t} | \bar{f}_t), \;\; \text{with} \;\; r^{\mathrm{OPC}}_t = f^{opt}_t - \tilde{f}^{\mathrm{OFC}}_t.
\end{equation}

We observed that in some cases neither of the above modes can outperform directly copying/propagating the respective block in reconstructed features at time $t-1$ such that $\hat{f}_{t} = \hat{f}_{t-1}$. Hence, this \textit{Feature Propagation} is adopted as the third prediction type described as:
\begin{equation}
    \tilde{f}^{\mathrm{FPG}}_t = \hat{f}_{t-1}, \;\; \text{with} \;\; r^{\mathrm{FPG}}_t = f^{opt}_t - \tilde{f}^{\mathrm{FPG}}_t.
\end{equation}

\begin{figure*}[h]
    \centering
    \includegraphics[width=0.95\linewidth]{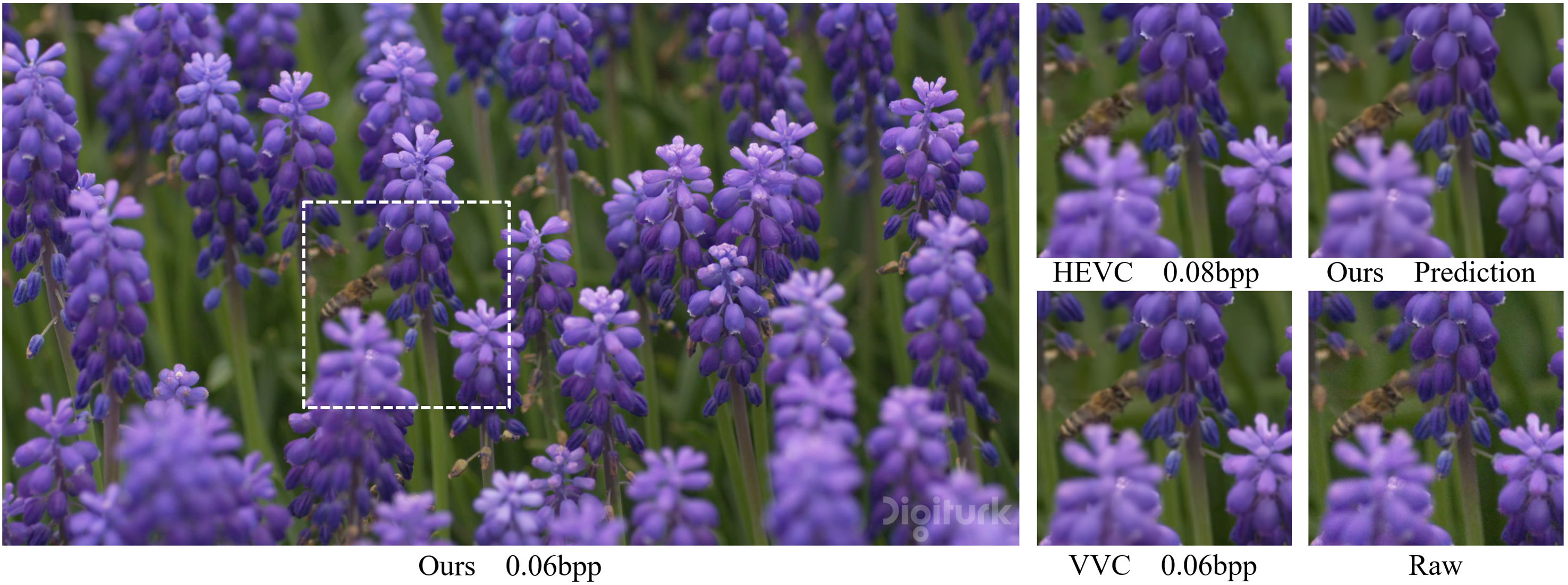}
    \vspace{-5pt}
    \caption{Reconstruction with  standard codecs (HEVC, VVC) and our MMVC method. Details of the static background and dynamic objects are well preserved in the frame generated from the predicted features and entropy-coded residuals that are block wise selected from multiple modes. Compared with HEVC, our result yields less block artifacts  preserving finer details. Our method achieves high quality similar to that of VVC codec, which is the state-of-the-art standard released recently.}
    \label{fig:bee}
    \vspace{-12pt}
\end{figure*}

After having the predicted feature representations under the above modes for all non-skip blocks, the residuals ($r^{\mathrm{FP}}_t, r^{\mathrm{OFC}}_t, r^{\mathrm{FPG}}_t$) are partitioned to equal-sized residual blocks ($r^{\mathrm{FP},i,j}_t, r^{\mathrm{OFC},i,j}_t, r^{\mathrm{FPG},i,j}_t$) so that each block  indexed by $i$ and $j$ has a set of residuals from different modes. The residual block partition side is determined to keep the number of blocks unchanged from  that of pixel domain blocks (i.e., $k \times k$). To determine the optimal prediction mode, we quantize and entropy encode (introduced in Section \ref{method-entropy}) each of the block partitioned residuals respectively, and proceed with the one that has the shortest code length. Therefore, the output of this step is a block-based residual map $r^{i,j}_t$ constructed by block wise optimal prediction mode selection. This process is described as:
\begin{equation}
\begin{split}
    & r_t^{i,j} = \argmin_{\hat{r}_t^{i,j}} (R(Q(\hat{r}_t^{i,j}))), \;\; \text{with} \;\; \\
    & \hat{r}^{i,j}_t \in \{r^{\mathrm{FP}\:i,j}_t, r^{\mathrm{OFC}\:i,j}_t, r^{\mathrm{FPG}\:i,j}_t\} \;\; \text{and} \;\; i,j \in \{1,\cdots,k\},
\end{split}
\end{equation}
where $Q(\cdot)$ and $R(\cdot)$ represent the quantization step and the bitrate after entropy coding respectively.

\subsection{Block Wise Channel Removal} 
\label{method-channel}

We adopt an adaptive residual channel removal technique to ensure that more bits are allocated to quality-critical residual elements. Carefully designed channel removal criteria can guarantee the reconstruction quality while reducing the number of bits consumed by unimportant residual feature channels. In an effort to only preserve feature channels carrying essential residuals and maintain the reconstruction quality after channel removal, we inspect each channel in a block separately. For one residual block and the predicted block along with it, the least important channel is selected by evaluating the PSNR degradation per channel removal. Channels are evaluated and removed iteratively in this manner as long as the quality degradation is within a predefined limit.



\subsection{Density-Adaptive Entropy Coding} \label{method-entropy}
The quantization and entropy coding is highlighted with the blue background in Figure \ref{fig:mode}. This datapath operates in accordance with the adaptive channel removal strategy, where pruned residual channels are set to zeros. It also considers sparse non-zero residuals as a result of efficient prediction. Our density-adaptive entropy coding method consists of a block wise sparse path and a dense path as shown in Figure \ref{fig:mode}. The density of each non-zero residual block is first evaluated, and the block is fed into the sparse path when the density is under a predefined threshold. Otherwise, the block is fed into the dense path. This mode selection is recorded as a block wise binary density map. 

In the sparse residual path, the non-zero residual positions are run-length coded prior to conventional arithmetic encoding, and non-zero residuals are gathered together for separate arithmetic encoding. The dense path consists of two options: (1) a learned entropy codec model guided by the hyperpriors followed by direct quantization \cite{NIPS2018_8275}, and (2) a conventional arithmetic coding method coupled with ADMM \cite{ADMM} quantization trained to non-uniformly discretize the residuals and optimized for minimal quantization error. We proceed with the option that leads to a lower rate and record the corresponding block wise entropy coding mode map $w_{t}$ for the receiver as side information.

Note that we maintain the same entropy coding path across all channels in each block to limit the cost of bits to encode $w_{t}$. To further reduce the bitrate, the binary density map (sparse vs. dense path) is also entropy coded with Huffman coding. This method incurs additional bits for conveying the side information (density map and $w_{t}$), our experiments in Section \ref{experiment-ablation} confirm that the overall bitrate reduction offsets the side information overhead.

\begin{figure*}[h]
    \centering
    \includegraphics[width=0.95\linewidth]{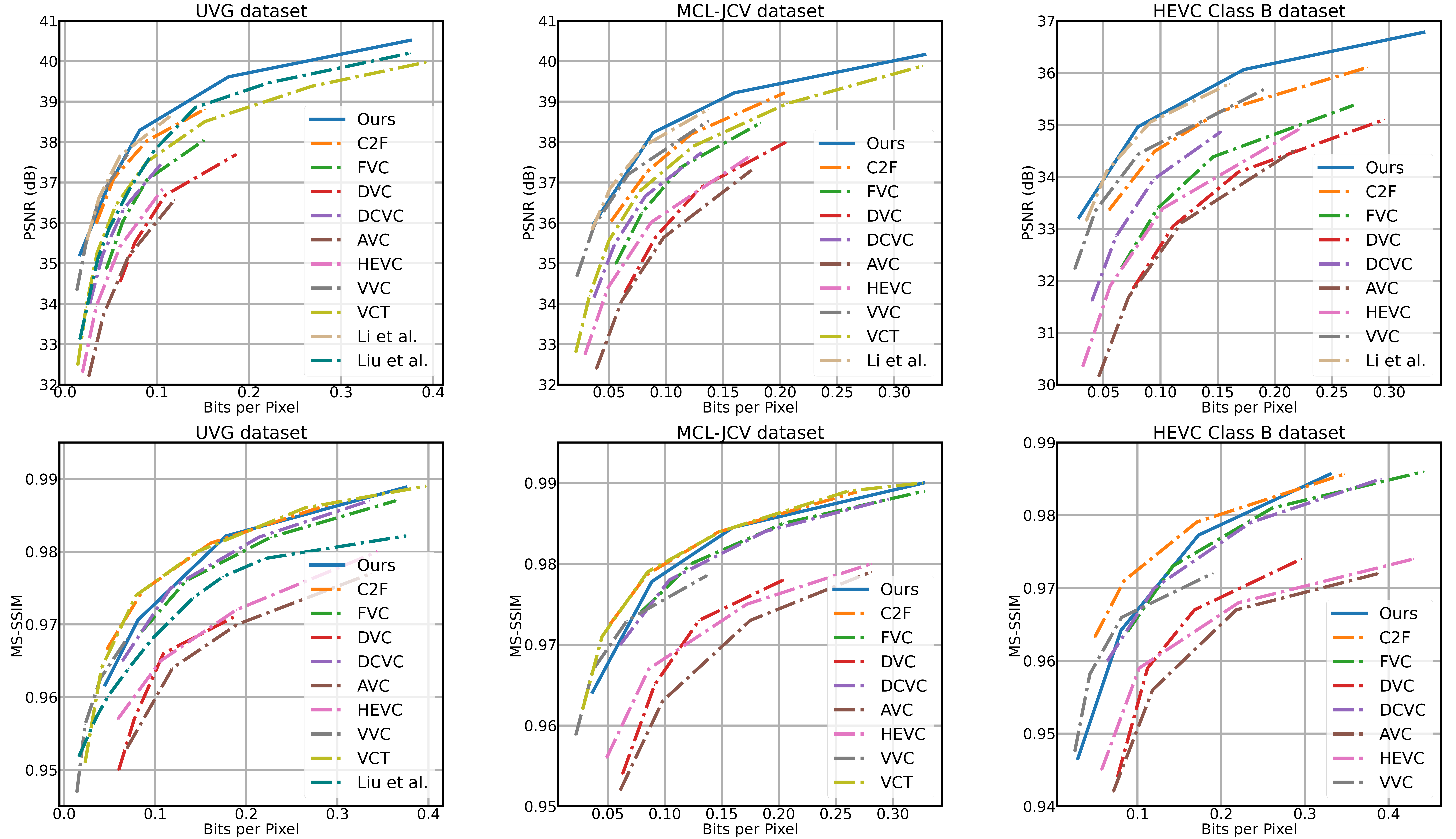}
    \vspace{-5pt}
    \caption{Rate-distortion curves measured on UVG, MCL-JCV, and HEVC Class B datasets in terms of PSNR and MS-SSIM.}
    \label{fig:rd_curves}
\end{figure*}

\begin{table*}[!]
\caption{Performance results evaluated by BD-Bitrate (BDBR) with PSNR metric (\%). We use VVC (\textit{low delay P} mode) as the anchor (i.e., BDBR = 0 for VVC). Negative values imply bitrate saving compared to VVC, while positive values imply the opposite. }
\vspace{-5pt}
\label{BD-Bitrate comparison}
\centering
\resizebox{\textwidth}{!}{
\begin{tabular}{ccccccccccccccc}
    \toprule
    \bf Dataset & \bf Ours & \bf  AVC & \bf  HEVC & \bf C2F &\bf  FVC &\bf  DVC & \bf DCVC & \bf VCT & \bf Li \textit{et al.} & \bf Liu \textit{et al.} & \bf HLVC & \bf M-LVC & \bf Agustsson \textit{et al.} & \bf RLVC \\
    \midrule
    \bf UVG & 0.81 & 246.70 & 171.98 & 12.59 & 112.14 & 210.75 & 98.19  & 65.49 & \bf-2.58 & 73.97 & 183.90 & 161.38 & 175.24 & 148.72 \\
    \bf MCL-JCV & \bf -16.73 & 188.00 & 124.56 & 19.44 & 72.45 & 150.08 & 67.71 & 44.92 & -11.25 & -- & -- & -- & 118.78 & --\\
    \bf HEVC Class B & \bf -28.28 & 198.79 & 127.10 & 14.59 & 108.01 & 176.82 & 66.14  & -- & -20.29 & -- & -- & 63.72 & -- & --\\
    \bottomrule
    \end{tabular}
    }
 \vspace{-12pt}
\end{table*}

\subsection{Model Training Strategy and Losses}
\label{method-losses}
Our mode selection scheme requires a uniform feature space for different prediction modes. Therefore both predictors ($P^{\mathrm{FP}}$ and $P^{\mathrm{OFC}}$) need to be optimized under the same pixel-feature space mapping. To get this mapping regularized under different rates, we first train our \textit{Optical Flow Conditioned Feature Prediction} model $P^{\mathrm{OFC}}$ optimized by:
\begin{equation}
\begin{split}
   \min_{\gamma, \eta, \phi, \varphi} R_{\gamma}(f^{opt}_t - {P^{\mathrm{OFC}}_{\eta}(E_{\phi}(\hat{x}_{<t})|E_{\phi}(\bar{x}_{t}))}) + \\ 
   \lambda \cdot d (D_{\varphi}(\tilde{f}^\mathrm{OFC}_t + \hat{r}^\mathrm{OFC}_t), x_t), 
\end{split}
\end{equation}
where $E(\cdot)$ and $D(\cdot)$ are the auto-encoder pair, and $d(\cdot)$ is the distortion calculated by MSE.

After this mapping is fixed (\textit{i.e.} the weights of encoder/decoder pair are trained) for $P^{\mathrm{OFC}}$ at a specific rate point, we then optimize the \textit{Feature Prediction} model $P^{\mathrm{FP}}$  to minimize distortion between the predicted and optimal features, measured by both MSE and discriminator loss. This optimization process is expressed as:
\begin{equation}
    \begin{split}
    \min _{\theta} \max _{\psi}\; (1-\alpha)\cdot \{\mathbb{E}_{f \sim p_{\text {opt}}(f)}[\log \textit{S}_{\psi}(D({f_t^{opt}}))] + \\
    \mathbb{E}_{f \sim p_{{f}}(f)}[\log (1-\textit{S}_{\psi}(D(P^{\mathrm{FP}}_{\theta}(\hat{f}_{<t}))))]\} + \\
    \alpha\cdot d(D(P^{\mathrm{FP}}_{\theta}(\hat{f}_{<t})), x_t),
    \end{split}
\end{equation}
where $\textit{S}(\cdot)$ is the discriminator network optimized in the GAN setting together with the prediction model to judge whether the reconstructed frame from the feature set is original (i.e., raw frame) or not. This discriminator model makes the training of the prediction model converge faster.  %

\begin{figure*}[h]
    \centering
    \includegraphics[width=0.95\linewidth]{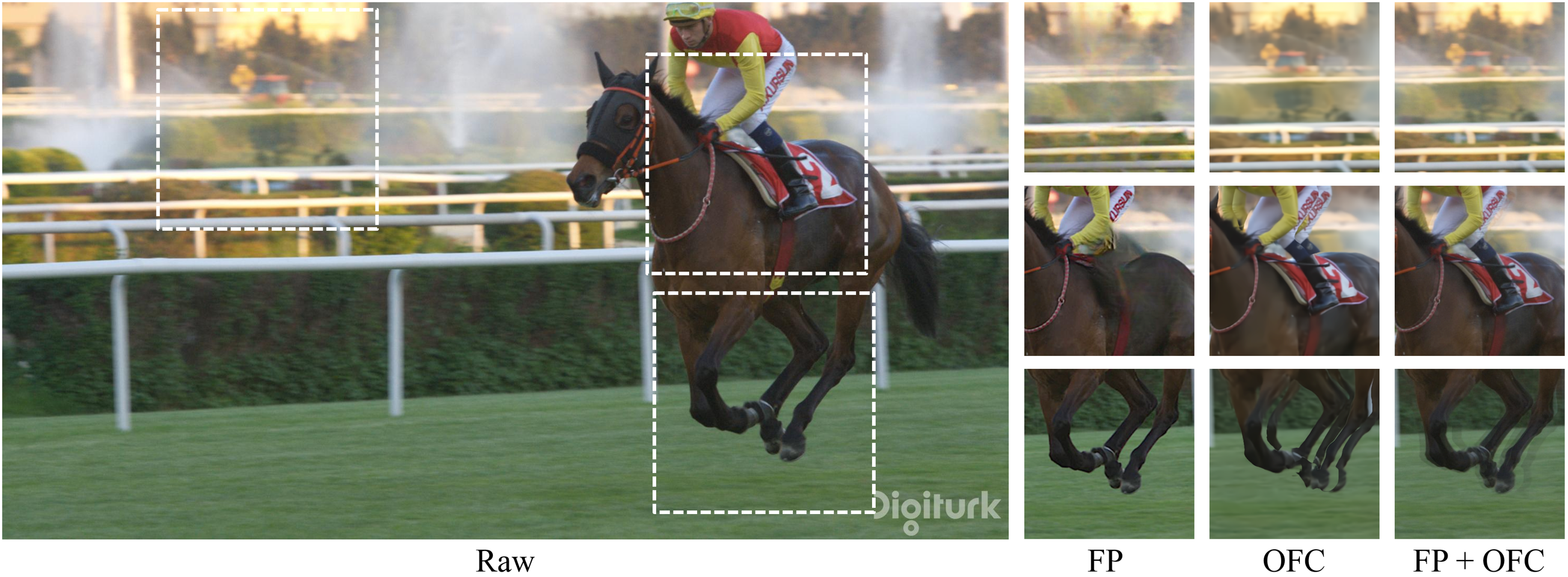}
    \vspace{-5pt}
    \caption{Reconstruction of multiple prediction modes without using the information from residual, where \textbf{FP} and \textbf{OFC} stands for \textit{Feature Prediction} and \textit{Optical Flow conditioned Feature Prediction} respectively.}
    \label{fig:pred_only}
    \vspace{-12pt}
\end{figure*}

\section{Experiments}
\subsection{Experimental Setup}
\textbf{Training datasets:} We use the  Vimeo-90k dataset,  Kinetics dataset, and UGC dataset for training purposes. The \textit{Optical Flow Conditioned Feature Prediction} mode is trained with the Vimeo-90k Septuplet \cite{xue2019video}, which contains 89,800 short video sequences, with each sequence having 7 consecutive frames of size $488\times 256$ pixels. To enlarge the training set, we randomly crop each original sequence to four $256\times 256$-pixel aggregated sequences. The \textit{Feature Prediction} mode is trained with part of the Kinetics dataset and the UGC dataset. The Kinetics dataset has 98,000 videos, each of 10 seconds with a resolution higher than 720p. The UGC dataset is composed of clips each lasting for 20 seconds. We combine the videos in Kinetics and UGC that have resolutions higher than 1080p, and crop the frames to $1024 \times 1024$ pixels for the training process.

\textbf{Testing datasets:} To evaluate the performance of our method quantitatively and qualitatively, we perform experiments on three datasets: the UVG dataset \cite{10.1145/3339825.3394937}, the MCL-JCV dataset \cite{MCL-JCV} and the HEVC class B dataset \cite{6316136}. All testing videos we choose have the same 1080p resolution. And to showcase the benefit of the \textit{Optical Flow Conditioned Feature Prediction} mode, we also adopt part of  the Kinetics dataset for ablative experiments. Video frames used for testing are not included in the training dataset.

\textbf{Evaluation metrics:} PSNR and MS-SSIM are used as the quantitative evaluation metrics in our experiments. PSNR is a standard way to reflect the degree of distortion in reconstruction whereas MS-SSIM often serves as a proxy indicator for perceptual quality.

\textbf{Training details:} We train the two prediction modes separately so that the training procedure can be divided into two stages. First, we train the encoder-decoder pair, the context and entropy model, together with the \textit{Optical Flow Conditioned Feature Prediction} model $P^{\mathrm{OFC}}$ (Feature Predictor path is disabled) end-to-end till convergence. We optimize our model involved in the experiments only with MSE as the distortion loss. To achieve different bitrates for rate-distortion tradeoffs, we curate a set of Lagrange multipliers as $\lambda = \{2, 64, 256, 1024, 2048, 4096\}$. The model is optimized for 10M steps with a batch size of 16. The learning rate is initialized to be $10^{-4}$, which is scaled to half every 2M steps. As each individual value $\lambda_i$ leads to a unique bitrate along with the quality, we end up having a set of trained encoder-decoder pairs for various rates. During the second stage in the training procedure, we fix the parameters in the trained encoder-decoder pair, and we proceed to obtain the optimal feature representations $f^{opt}_t$ from iterative back-propagation through the decoder. These obtained optimal features serve as the input to train the \textit{Feature Prediction} model $P^{\mathrm{FP}}$, which is optimized by discriminator loss in addition to the MSE loss for enhanced visual quality. We train $P^{\mathrm{FP}}$ model at an initial learning rate of $5\times10^{-4}$ for 20M steps with a batch size of 8, and we decay the learning rate by half for every 2M steps after training 10M steps.


\subsection{Results and Analysis}
\textbf{Quantitative Results:} To demonstrate the performance of our proposed MMVC, we evaluate our rate and distortion tradeoff curves and PSNR-based BD-Bitrate measurements with the state-of-the-art learned video compression algorithms published in recent years. Specifically, we include results from DVC \cite{Lu_2019_CVPR}, FVC \cite{Hu_2021_CVPR}, Liu \textit{et al.} \cite{Liu_2021_CVPR}, DCVC \cite{li2021deep}, C2F \cite{Hu_2022_CVPR}, VCT \cite{VCT}, Li \textit{et al.} \cite{Li_2022_HST} \cite{Li_2022_HST}, HLVC \cite{yang2020learning}, M-LVC \cite{lin2020m}, Agustsson \textit{et al.} \cite{agustsson2020scale}, and RLVC \cite{yang2021learning}.  
We also include measurements from traditional codec standards: AVC \cite{1218189}, HEVC \cite{6316136}, and the latest VVC \cite{9301847}. Following the prior work \cite{li2021deep}, we choose to encode AVC and HEVC under \textit{veryslow} mode with a GoP of 12/10. Compared with \textit{veryfast}, the \textit{veryslow} mode compresses video frames to a lower bitrate at the cost of longer encoding time, which aligns better with our target to generate high quality frames with the lowest bitrate but potentially longer latency. For VVC encoding, we use the \textit{low delay P} mode with a GoP size of 100 and set the IntraPeriod to be 4. 

Figure \ref{fig:rd_curves} presents the rate-distortion curves measured with PSNR and MS-SSIM on UVG, MCL-JCV, and HEVC Class B datasets. The performance curves demonstrate that our proposed method outperforms the state-of-the-art learning-based approaches and conventional codecs in terms of PSNR for most of the bitrates we cover. Particularly at 0.1 bit-per-pixel (bpp), our approach achieves 2dB quality improvement on average compared with HEVC (\textit{veryslow}) for all testing datasets. Although not specially trained or fine-tuned for MS-SSIM, our method achieves comparative performance under the MS-SSIM metric consistently across all test datasets, especially in higher bpp regions. Table \ref{BD-Bitrate comparison} shows the BD-Bitrate (BDBR) results in terms of PSNR anchored to VVC. The evaluation is based on UVG, MCL-JCV, and HEVC Class B datasets. Our method demonstrates competitive or superior performance compared to other schemes in Table \ref{BD-Bitrate comparison}.

\begin{table*}
\caption{Mode utilization and performance on the UVG dataset and Kinetics dataset, where FP, OFC, FPG, and S stands for \textit{Feature Prediction}, \textit{Optical Flow Conditioned Feature Prediction}, \textit{Feature Propagation}, and \textit{Skip} mode respectively.}
\label{uvg_kinetics}
\vspace{-5pt}
\begin{center}
\begin{tabular}{c|cccc|cccc}
    \toprule
    \bf Dataset & \multicolumn{4}{c|}{\bf UVG} & \multicolumn{4}{c}{\bf Kinetics} \\
    \hline
    \multirow{2}{*}{\bf Prediction mode} &\multirow{2}{*}{\bf PSNR (dB)} &\bf Removed  &\multirow{2}{*}{\bf Bpp} &\bf Bitrate &\multirow{2}{*}{\bf PSNR (dB)}  &\bf Removed  &\multirow{2}{*}{\bf Bpp} &\bf Bitrate \\
     &  &\bf channels  &  & \bf saving &  &\bf channels  &  & \bf saving \\
    \hline
     FP  & 38.0  & 23\% & 0.146 & 0\% & 37.7  & 29.8\% & 0.136 & 0\%  \\
     OFC  & 36.9  & 47\% & 0.118 & 19.2\%  & 37.4  & 49.3\% & 0.106 & 22.1\% \\
     FP+OFC   & 38.1  &27\% &0.096 & 34.3\% & 37.7  &41.6\% &0.099 & 27.2\% \\
     FP+OFC+FPG  & 38.2  & 27\% & 0.084 & 42.5\% & 37.7  & 43.5\% & 0.096 & 29.4\% \\
     FP+OFC+FPG+S  & 38.2  & 44\% & 0.081  &44.5\% & 37.8  & 50.8\% & 0.088  &35.3\% \\
    \hline
    \multirow{2}{*}{\bf Mode utilization} & \bf FP & \bf OFC & \bf FPG & \bf S & \bf FP & \bf OFC & \bf FPG & \bf S \\
     & 78.1\% & 10.6\% & 6\% & 5.3\% & 37.6\% & 38.3\% & 12\% & 11.6\% \\
    \bottomrule
    \end{tabular}
\end{center}
\vspace{-12pt}
\end{table*}


\textbf{Qualitative Analysis:}
Figure \ref{fig:bee} shows example reconstructed frames from the UVG dataset. Our approach exhibits similar quality (if not more visually appealing) with a bpp comparable to other codecs. The details of dynamic motions are well preserved at less than $0.1$ bpp, demonstrating that MMVC can accurately predict the next frame by a combination of different modes. For the background field, the sharpness of our reconstructed frame is (subjectively) better than other standard codecs. The complementary nature of different prediction modes in our scheme is visualized in Figure \ref{fig:pred_only}, which shows the decoded scenes directly obtained from the predicted features using a specific mode for the entire frame without residual compensation. Using FP mode only leads to sharper details in general, but it loses some contents such as the vehicle behind the fountain and the rider's leg. On the contrary, applying OFC mode only results in unfaithful reconstructions near the horse's legs. By adopting multiple prediction modes that complement each other (FP + OFC), our prediction is able to cover content variety in the original frame. The resulting residual is sparse and can be condensed to a shorter bitstream.

\subsection{Ablation Study}
\label{experiment-ablation}
\textbf{Mode utilization:} We summarize mode utilization for the UVG dataset and 20 selected video sequences from the Kinetics dataset at relatively low bitrates of 0.081 and 0.088, respectively. As presented in Table \ref{uvg_kinetics}, this analysis involves evaluating the separate performance of each prediction mode, examining the impact of utilizing multiple prediction modes, and quantifying gains provided by skipping the encoding of unchanged blocks (\textit{i.e}., \textit{Skip} mode). 

As shown in  Table \ref{uvg_kinetics}, the \textit{Feature Prediction} and \textit{Optical Flow Conditioned Feature Prediction} modes achieve comparable performance. For UVG, FP slightly outperforms OFC with 1dB higher PSNR and only 20\% higher bitrate. Meanwhile, OFC is more favorable than FP for the Kinetics dataset. By adopting the ensemble of both modes (FP + OFC), the quality is preserved with an even lower bitrate, indicating that these two prediction modes can complement each other by capturing different motion patterns. Including the \textit{Feature Propagation} mode as an alternative prediction path further reduces the bitrate without degrading the quality. The compression ratio improves noticeably by introducing the \textit{Skip} mode as the final additional mode. For sequences in the UVG dataset, we observe that usage of FP surpasses other modes significantly. However, in the Kinetics dataset where the scenes are captured mostly by a fixed-view camera showcases higher utilization of the other modes. Fixed backgrounds in Kinetics sequences enable higher utilization of the \textit{Skip} mode for significant bitrate reduction. In general, introducing \textit{Skip} mode reduces the required bitrate for the same quality. 

\begin{table}[t]
\caption{Percentage of additional bitrate saving from density-adaptive entropy coding module compared to the baseline of using FP mode and dense path only on the Kinetics dataset. Note that the S mode is not included because it does not involve any residual coding.}
\vspace{-5pt}
\label{entropy_mode}
\centering
\resizebox{\columnwidth}{!}{
\begin{tabular}{cccc}
    \toprule
    \small \bf Prediction mode   & \small \bf Dense path only & \small \bf Dense + sparse paths \\
    \midrule
    FP   & 0\%  & 4.1\% \\ 
    FP+OFC  & 6.3\% & 21.2\%\\
    FP+OFC+FPG  & 6.5\% &23.9\%\\
    \bottomrule
    \end{tabular}}
    \vspace{-12pt}
\end{table}

\textbf{Effectiveness of channel removal and density-adaptive entropy coding:} One column in Table \ref{uvg_kinetics} shows the percentage of removed residual channels for various prediction modes. It shows that the percentage of removed residual channels is generally higher as the prediction becomes more accurate with mode selections from the full ensemble of available prediction modes (FP+OFC+FPG+S) compared to the single mode case (FP only). 

Table \ref{entropy_mode} summarizes the additional bitrate saving by the density-adaptive entropy coding compared to the baseline of using FP mode and dense-path only. The evaluation is based on the Kinetics dataset at a relatively low bpp of 0.165, where the utilization of each prediction mode is well balanced. Allowing more prediction modes generally reduces the density of the residual. Thus the proposed density-adaptive entropy coding provides more significant savings (an additional 23.9\% saving) when it is combined with the full ensemble of available prediction modes (FP+OFC+FPG). This saving includes the overhead of sending the density map and mode selection side information. 
\section{Conclusion}
In this work, we present a dynamic mode selection based video coding scheme MMVC. It can dynamically switch between multiple prediction paths adapting to distinct motion patterns that appear on different blocks within a frame. To further reduce the required bitrate for the prediction residual encoding, we propose a channel removal approach together with a density-adaptive entropy coding scheme to attain more compact residual representations when the residual entropy and density significantly vary block wise. Evaluations with various test datasets confirm that our method can attain outstanding rate-distortion trade-offs.

{\small
\bibliographystyle{ieee_fullname}
\bibliography{ref}

\begin{thebibliography}{10}\itemsep=-1pt

\bibitem{Agustsson_2020_CVPR}
Eirikur Agustsson, David Minnen, Nick Johnston, Johannes Balle, Sung~Jin Hwang,
  and George Toderici.
\newblock Scale-space flow for end-to-end optimized video compression.
\newblock In {\em Proceedings of the IEEE/CVF Conference on Computer Vision and
  Pattern Recognition (CVPR)}, June 2020.

\bibitem{agustsson2020scale}
Eirikur Agustsson, David Minnen, Nick Johnston, Johannes Balle, Sung~Jin Hwang,
  and George Toderici.
\newblock Scale-space flow for end-to-end optimized video compression.
\newblock In {\em Proceedings of the IEEE/CVF Conference on Computer Vision and
  Pattern Recognition}, pages 8503--8512, 2020.

\bibitem{balle2017endtoend}
Johannes Ball{\'e}, Valero Laparra, and Eero~P. Simoncelli.
\newblock End-to-end optimized image compression.
\newblock In {\em International Conference on Learning Representations}, 2017.

\bibitem{balle2018variational}
Johannes Ballé, David Minnen, Saurabh Singh, Sung~Jin Hwang, and Nick
  Johnston.
\newblock Variational image compression with a scale hyperprior.
\newblock In {\em International Conference on Learning Representations}, 2018.

\bibitem{Cui_2021_CVPR}
Ze Cui, Jing Wang, Shangyin Gao, Tiansheng Guo, Yihui Feng, and Bo Bai.
\newblock Asymmetric gained deep image compression with continuous rate
  adaptation.
\newblock In {\em Proceedings of the IEEE/CVF Conference on Computer Vision and
  Pattern Recognition (CVPR)}, pages 10532--10541, June 2021.

\bibitem{8237351}
Jifeng Dai, Haozhi Qi, Yuwen Xiong, Yi Li, Guodong Zhang, Han Hu, and Yichen
  Wei.
\newblock Deformable convolutional networks.
\newblock In {\em 2017 IEEE International Conference on Computer Vision
  (ICCV)}, pages 764--773, 2017.

\bibitem{Habibian_2019}
Amirhossein Habibian, Ties~Van Rozendaal, Jakub Tomczak, and Taco Cohen.
\newblock Video compression with rate-distortion autoencoders.
\newblock In {\em 2019 {IEEE}/{CVF} International Conference on Computer Vision
  ({ICCV})}. {IEEE}, oct 2019.

\bibitem{Hu_2022_CVPR}
Zhihao Hu, Guo Lu, Jinyang Guo, Shan Liu, Wei Jiang, and Dong Xu.
\newblock Coarse-to-fine deep video coding with hyperprior-guided mode
  prediction.
\newblock In {\em Proceedings of the IEEE/CVF Conference on Computer Vision and
  Pattern Recognition (CVPR)}, pages 5921--5930, June 2022.

\bibitem{Hu_2021_CVPR}
Zhihao Hu, Guo Lu, and Dong Xu.
\newblock Fvc: A new framework towards deep video compression in feature space.
\newblock In {\em Proceedings of the IEEE/CVF Conference on Computer Vision and
  Pattern Recognition (CVPR)}, pages 1502--1511, June 2021.

\bibitem{Ladune_2020_ModeNet}
Théo Ladune, Pierrick Philippe, Wassim Hamidouche, Lu Zhang, and Olivier
  Déforges.
\newblock Modenet: Mode selection network for learned video coding.
\newblock In {\em 2020 IEEE 30th International Workshop on Machine Learning for
  Signal Processing (MLSP)}, pages 1--6, 2020.

\bibitem{li2021deep}
Jiahao Li, Bin Li, and Yan Lu.
\newblock Deep contextual video compression.
\newblock In A. Beygelzimer, Y. Dauphin, P. Liang, and J.~Wortman Vaughan,
  editors, {\em Advances in Neural Information Processing Systems}, 2021.

\bibitem{Li_2022_HST}
Jiahao Li, Bin Li, and Yan Lu.
\newblock Hybrid spatial-temporal entropy modelling for neural video
  compression.
\newblock In {\em Proceedings of the 30th ACM International Conference on
  Multimedia}, MM '22, page 1503–1511, New York, NY, USA, 2022. Association
  for Computing Machinery.

\bibitem{lin2020m}
Jianping Lin, Dong Liu, Houqiang Li, and Feng Wu.
\newblock M-lvc: Multiple frames prediction for learned video compression.
\newblock In {\em Proceedings of the IEEE/CVF Conference on Computer Vision and
  Pattern Recognition}, pages 3546--3554, 2020.

\bibitem{liu_2020_BPGAN}
B. {Liu}, A. {Cao}, and H. {Kim}.
\newblock Unified signal compression using generative adversarial networks.
\newblock In {\em ICASSP 2020 - 2020 IEEE International Conference on
  Acoustics, Speech and Signal Processing (ICASSP)}, pages 3177--3181, 2020.

\bibitem{Liu_2021_CVPR}
Bowen Liu, Yu Chen, Shiyu Liu, and Hun-Seok Kim.
\newblock Deep learning in latent space for video prediction and compression.
\newblock In {\em Proceedings of the IEEE/CVF Conference on Computer Vision and
  Pattern Recognition (CVPR)}, pages 701--710, June 2021.

\bibitem{Lu_2019_CVPR}
Guo Lu, Wanli Ouyang, Dong Xu, Xiaoyun Zhang, Chunlei Cai, and Zhiyong Gao.
\newblock Dvc: An end-to-end deep video compression framework.
\newblock In {\em The IEEE Conference on Computer Vision and Pattern
  Recognition (CVPR)}, June 2019.

\bibitem{VCT}
Fabian Mentzer, George Toderici, David Minnen, Sung-Jin Hwang, Sergi Caelles,
  Mario Lucic, and Eirikur Agustsson.
\newblock Vct: A video compression transformer, 2022.

\bibitem{10.1145/3339825.3394937}
Alexandre Mercat, Marko Viitanen, and Jarno Vanne.
\newblock Uvg dataset: 50/120fps 4k sequences for video codec analysis and
  development.
\newblock In {\em Proceedings of the 11th ACM Multimedia Systems Conference},
  MMSys '20, page 297–302, New York, NY, USA, 2020. Association for Computing
  Machinery.

\bibitem{NIPS2018_8275}
David Minnen, Johannes Ball\'{e}, and George~D Toderici.
\newblock Joint autoregressive and hierarchical priors for learned image
  compression.
\newblock In S. Bengio, H. Wallach, H. Larochelle, K. Grauman, N. Cesa-Bianchi,
  and R. Garnett, editors, {\em Advances in Neural Information Processing
  Systems 31}, pages 10771--10780. Curran Associates, Inc., 2018.

\bibitem{ADMM}
Ao Ren, Tianyun Zhang, Shaokai Ye, Jiayu Li, Wenyao Xu, Xuehai Qian, Xue Lin,
  and Yanzhi Wang.
\newblock {ADMM-NN:} an algorithm-hardware co-design framework of dnns using
  alternating direction method of multipliers.
\newblock {\em CoRR}, abs/1812.11677, 2018.

\bibitem{Rippel_2019_ICCV}
Oren Rippel, Sanjay Nair, Carissa Lew, Steve Branson, Alexander~G. Anderson,
  and Lubomir Bourdev.
\newblock Learned video compression.
\newblock In {\em Proceedings of the IEEE/CVF International Conference on
  Computer Vision (ICCV)}, October 2019.

\bibitem{9301847}
Gary Sullivan.
\newblock Versatile video coding (vvc) arrives.
\newblock In {\em 2020 IEEE International Conference on Visual Communications
  and Image Processing (VCIP)}, pages 1--1, 2020.

\bibitem{6316136}
Gary~J. Sullivan, Jens-Rainer Ohm, Woo-Jin Han, and Thomas Wiegand.
\newblock Overview of the high efficiency video coding (hevc) standard.
\newblock {\em IEEE Transactions on Circuits and Systems for Video Technology},
  22(12):1649--1668, 2012.

\bibitem{teed2020RAFT}
Zachary Teed and Jia Deng.
\newblock {RAFT:} recurrent all-pairs field transforms for optical flow.
\newblock In {\em European Conference on Computer Vision}, 2020.

\bibitem{MCL-JCV}
Haiqiang Wang, Weihao Gan, Sudeng Hu, Joe~Yuchieh Lin, Lina Jin, Longguang
  Song, Ping Wang, Ioannis Katsavounidis, Anne Aaron, and C.-C.~Jay Kuo.
\newblock Mcl-jcv: A jnd-based h.264/avc video quality assessment dataset.
\newblock In {\em 2016 IEEE International Conference on Image Processing
  (ICIP)}, pages 1509--1513, 2016.

\bibitem{1218189}
T. Wiegand, G.J. Sullivan, G. Bjontegaard, and A. Luthra.
\newblock Overview of the h.264/avc video coding standard.
\newblock {\em IEEE Transactions on Circuits and Systems for Video Technology},
  13(7):560--576, 2003.

\bibitem{wu2018vcii}
Chao-Yuan Wu, Nayan Singhal, and Philipp Kr{\"a}henb{\"u}hl.
\newblock Video compression through image interpolation.
\newblock In {\em ECCV}, 2018.

\bibitem{xue2019video}
Tianfan Xue, Baian Chen, Jiajun Wu, Donglai Wei, and William~T Freeman.
\newblock Video enhancement with task-oriented flow.
\newblock {\em International Journal of Computer Vision (IJCV)},
  127(8):1106--1125, 2019.

\bibitem{yang2020learning}
Ren Yang, Fabian Mentzer, Luc~Van Gool, and Radu Timofte.
\newblock Learning for video compression with hierarchical quality and
  recurrent enhancement.
\newblock In {\em Proceedings of the IEEE/CVF Conference on Computer Vision and
  Pattern Recognition}, pages 6628--6637, 2020.

\bibitem{yang2021learning}
Ren Yang, Fabian Mentzer, Luc Van~Gool, and Radu Timofte.
\newblock Learning for video compression with recurrent auto-encoder and
  recurrent probability model.
\newblock {\em IEEE Journal of Selected Topics in Signal Processing},
  15(2):388--401, 2021.

\bibitem{yang2021hierarchical}
Ruihan Yang, Yibo Yang, Joseph Marino, and Stephan Mandt.
\newblock Hierarchical autoregressive modeling for neural video compression.
\newblock In {\em International Conference on Learning Representations}, 2021.

\end{thebibliography}
}

\end{document}